\documentclass[12pt]{article}
\usepackage{epsf}

\newcommand{\cZ}{{\cal Z}}
\newcommand{\cO}{{\cal O}}
\newcommand{\tcO}{{\tilde {\cal O}}_{\mathrm{MaA}}}

\newcommand{\dD}{{\cal D}}

\newcommand{\eq}[1]{(\ref{#1})}
\newcommand{\dual}{\mbox{}^{\ast}}
\newcommand{\diff}{\partial}

\newcommand{\beq}{\begin{equation}}
\newcommand{\eeq}{\end{equation}}
\newcommand{\beqn}{\begin{eqnarray}}
\newcommand{\eeqn}{\end{eqnarray}}

\newcommand{\dd}{\mbox{d}}

\newcommand{\abstracts}[1]{{
\centering{\begin{minipage}{12.2truecm}
\normalsize\baselineskip=15pt
\centerline{\footnotesize ABSTRACT}\vspace*{0.3cm}
\parindent=20pt #1
\end{minipage}}\par}}

\begin{document}

\begin{flushright}
{\large ITEP-TH-76/98 \\
\vskip 1mm
FISIST/18-98/CFIF}
\end{flushright}
\vspace{1.5cm}

\begin{center}

{\baselineskip=24pt
{\Large \bf Effective Monopole Potential\\
for $\mathbf{SU(2)}$ Lattice Gluodynamics\\
in Spatial Maximal Abelian Gauge}\\}

{\baselineskip=16pt

\vspace{1cm}

{\large M.~N.~Chernodub, M.~I.~Polikarpov and A.I.~Veselov}

\vspace{.5cm}

{\large \it

Institute of Theoretical and Experimental Physics\\
B.~Cheremushkinskaya 25, Moscow, 117259, Russia}}
\end{center}

\vspace{1cm}

\abstracts{We investigate the dual superconductor hypothesis in
finite-temperature $SU(2)$ lattice gluodynamics in the {\it
Spatial} Maximal Abelian gauge. This gauge is more physical than
the ordinary Maximal Abelian gauge due to absence of
non-localities in temporal direction.  We show numerically that in
the Spatial Maximal Abelian gauge the probability distribution of
the abelian monopole field is consistent with the dual
superconductor mechanism of confinement: the abelian condensate
vanishes in the deconfinement phase and is not zero in the
confinement phase.}

\newpage

The dual superconductor hypothesis of color confinement~\cite{Man76}
in gluodynamics has been confirmed by various lattice
calculations~\cite{Reviews} in the so-called Maximal Abelian (MaA)
projection~\cite{KrScWi87}. This hypothesis is based on a partial gauge
fixing of a non-abelian group up to its abelian subgroup. After
gauge is fixed abelian monopoles arise due to singularities in the gauge
fixing conditions~\cite{tHo81}. If monopoles are condensed then the
vacuum of gluodynamics behaves as a dual superconductor and the electric
charges (quarks) in such a vacuum are confined.

The MaA projection on the lattice is defined by the
condition~\cite{KrScWi87}:
\beqn
\max_\Omega R_{\mathrm{MaA}}[U^\Omega]\,,\qquad
R_{\mathrm{MaA}}[U] = \sum\limits_l
{\mathrm{Tr}} [\sigma_3 U_l^+ \sigma_3 U_l]\,,
\quad l = \{x,\mu\}\,,
\label{MaA}
\eeqn
where the summation is over all lattice links and
$U_{x,\mu}$ are the lattice $SU(2)$ gauge fields.

The gauge fixing condition \eq{MaA} contains time components of the gauge
fields, $U_{x,4}$, therefore abelian operators in the MaA projection
correspond to nonlocal in time operators in terms of the original $SU(2)$
fields $U_{x,\mu}$. To show this let us consider the expectation value of
the $U(1)$ invariant operator $\cO$ in the Maximal Abelian
gauge~\cite{ChPoVe95,YukawaSchool}:
\beqn
{<\cO>}_{\mathrm{MaA}} =
\frac{1}{\cZ_{\mathrm{MaA}}} \int \dD U \, \exp\{ - S[U] + \lambda
R_{\mathrm{MaA}}[U]\} \,
\Delta_{FP}[U;\lambda] \, \cO(U)\,,\quad \lambda \to + \infty\,,
\label{E1}
\eeqn
where $\cZ_{\mathrm{MaA}} = < 1 >_{\mathrm{MaA}}$ is the partition
function in the fixed gauge.
$\Delta_{FP}[U;\lambda]$ is the Faddeev--Popov determinant:
\beqn
1=\Delta_{FP}[U;\lambda] \cdot \int \dD \Omega \, \exp \{ \lambda
R_{\mathrm{MaA}}[U^\Omega]\}\,, \quad \lambda \to + \infty\,.
\nonumber
\eeqn
Shifting the fields $U \to U^{\Omega^+}$ in eq.\eq{E1} and integrating
over $\Omega$ both in the nominator and denominator, we get:
\beqn
{<\cO>}_{\mathrm{MaA}} = <{\tcO}>\,,\quad
{\tcO}(U) = \frac{\int \dD \Omega \,
\exp\{ \lambda R_{\mathrm{MaA}}[U^\Omega] \} \, \cO(U^\Omega)}{\int \dD
\Omega \, \exp\{ \lambda R_{\mathrm{MaA}}[U^\Omega] \}}\,,
\nonumber
\eeqn
${\tcO}$ is the $SU(2)$ invariant operator.  Since
$\lambda\to + \infty$ we can use the saddle point approximation to calculate
${\tcO}$:
\beqn
{\tcO}(U) = \frac{\sum\limits^{N(U)}_{j=1}
Det^\frac{1}{2} M_{\mathrm{MaA}}[U^{\Omega^{(j)}}] \,
\cO(U^{\Omega^{(j)}})}{\sum\limits^{N(A)}_{k=1}
Det^\frac{1}{2} M_{\mathrm{MaA}}[U^{\Omega^{(k)}}]}\,,
\label{tO}
\eeqn
here $\Omega^{(j)}$ corresponds to the $N$--degenerate
global maxima of the functional $R_{\mathrm{MaA}}[U^\Omega]$
with respect to the regular gauge transformations $\Omega$:
$R_{\mathrm{MaA}}[U^{\Omega^{(j)}}] = R_{\mathrm{MaA}}[U^{\Omega^{(k)}}]$,
$j,k=1,\dots,N$. The matrix $M_{\mathrm{MaA}}$ is the Faddeev--Popov
operator~\cite{ChPoVe95}:
\beqn
M^{x,a;y,b}_{\mathrm{MaA}}[U] = \frac{\diff^2
R_{\mathrm{MaA}}(U^{\Omega(\omega)})}{\diff \omega^a (x) \,
\diff \omega^b (y)} {\lower0.15cm\hbox{${\Biggr |}_{\omega=0}$}}\,,
\nonumber
\eeqn
$\Omega(\omega) = \exp\{i\omega^a T^a\}$, $T^a=\sigma^a \slash 2$
are the generators of the gauge group, $\sigma^a$ are the Pauli matrices.

Since the gauge fixing functional $R_{\mathrm{MaA}}$ contains time
components of the gauge field $U$ then the operator ${\tilde \cO}(U)$ is
non-local in time. For time-nonlocal operators there are obvious
difficulties with the transition from the Euclidean to Minkowski
space--time.  Thus there are problems with physical interpretation of the
results obtained for abelian operators in the MaA projection.

The MaA gauge condition can be easily modified to overcome
this time-non-locality problem. The corresponding gauge condition is given
by:
\beqn
\max_\Omega R_{\mathrm{SMaA}}[\hat{U}^\Omega]\,,\qquad
R_{\mathrm{SMaA}}[U] = \sum\limits_{\mathbf{l}}
{\mathrm{Tr}}[\sigma_3 U_{\mathbf{l}}^+ \sigma_3 U_{\mathbf{l}}]\,,\quad
{\mathbf{l}}=\{x,i\}\,,\quad i = 1,2,3\,,
\label{SMaA}
\eeqn
where the summation is taken only over the spatial links ${\mathbf{l}}$.
We refer to this projection as the Spatial Maximal Abelian (SMaA)
projection\footnote{This gauge was discussed by U.-J.~Wiese in 1990, was
recently rediscovered by D.~Zwanzinger (private communication to M.I.P.),
and discussed by M.~M\"uller-Preussker at the 1997 Yukawa International
Seminar on {\it "Non-perturbative QCD - Structure of QCD Vacuum -"
(YKIS'97)}, 2-12 December, 1997, Yukawa Institute for Theoretical Physics,
Kyoto, Japan.}. In the SMaA gauge the gauge invariant operator \eq{tO} is
local in time.

In this paper we study the abelian monopole condensate $\Phi_c$
in SMaA projection. To calculate $\Phi_c$ we need the monopole
creation operator $\Phi_{\mathrm{mon}}(x)$. This operator was
found by Fr\"ohlich
and Marchetti~\cite{FrMa87} for compact electrodynamics
and was studied numerically in Refs.~\cite{Wiese}.
The Fr\"ohlich--Marchetti operator was generalized to the
abelian projection of lattice $SU(2)$ gluodynamics
in Refs.~\cite{OurPapers} where it was found that in the MaA projection
the monopole field is condensed in the confinement phase and $\Phi_c$
vanishes in the deconfinement phase\footnote{The similar results were
obtained for another definitions of $\Phi_{\mathrm{mon}}(x)$
\cite{DiGi,Nak96}.}.

The construction~\cite{OurPapers}
of the monopole creation operator for an arbitrary
Abelian projection is the following. We
parametrize the $SU(2)$ link matrix in the standard way:
$U^{11}_{x\mu} = \cos \phi_{x\mu}\, e^{i\theta_{x\mu}}; \
U^{12}_{x\mu} = \sin
\phi_{x\mu}\, e^{i\chi_{x\mu}};$ $\ U^{22}_{x\mu} = U^{11 *}_{x\mu}; \
U^{21}_{x\mu} = - U^{12 *}_{x\mu};$ $\ 0 \le \phi \le \pi/2, \ -\pi <
\theta,\chi \le \pi$.
The plaquette action in terms of the angles $\phi, \ \theta $ and
$\chi$ can be written as follows: $S_P  =  \frac{1}{2}\mbox{Tr}\, U_1 U_2
U_3^+ U_4^+ = S^a + S^n + S^i$, where
$S^a  = \cos \theta_P\, \cos\phi_1\, \cos\phi_2\, \cos\phi_3\,
\cos\phi_4$, $S^n$ and $S^i$ describe the interaction of the fields
$\theta$ and $\chi$
and self--interaction of the field $\chi$~\cite{ChPoVe95},
here the subscripts $1,...,4$ correspond to the links of the plaquette:  $1
\rightarrow \{x,x+\hat{\mu}\},...,4 \rightarrow \{x,x+\hat{\nu}$\}.
For a fixed abelian projection, each term $S^a$, $S^n$ and $S^i$ is
invariant under the residual $U(1)$ gauge transformations:
$\theta_{x\mu} \to \theta_{x\mu} +\alpha_x -\alpha_{x+\hat{\mu}}$,
$\chi_{x\mu} \to \chi_{x\mu} +\alpha_x + \alpha_{x+\hat{\mu}}$.

The operator $\Phi_{\mathrm{mon}}(x)$ creates the monopole
at the point $x$ on the dual lattice with a cloud of dual photons,
it is defined as follows~\cite{OurPapers}:
\beqn
   \Phi_{\mathrm{mon}}(x)
   = \exp \left\{ \sum_P \tilde\beta \left[ - \cos(\theta_P) +
   \cos(\theta_P + W_P(x)) \right] \right\}\,, \label{Ux2}
\eeqn
where $\tilde\beta = \beta \,
\cos\phi_1 \cos\phi_2 \cos\phi_3\cos\phi_4$,
$W_P$ is defined as follows~\cite{FrMa87}: $W_P
= 2 \pi \delta\Delta^{-1}(D_x-\omega_x))$. The integer valued 1-form
$\dual \omega_x$ represents the Dirac string attached to the monopole
\cite{FrMa87} and satisfies the equation: $\delta \dual \omega_x = \dual
\delta_x$. The function $D_x=\dd_{(3)} \Delta^{-1}_{(3)} \delta_x$
represents the cloud of dual photons.
Here $\dual \delta_x$ is the lattice $\delta$--function, it equals to
unity at the site $x$ of the dual lattice and is zero at the other sites;
$\dd_{(x)}$ and $\Delta^{-1}_{(3)}$ are the lattice derivative and
the inverse Laplace operator on three-dimensional time-slice which
includes the point $x$.

We study the monopole creation operator $\Phi_{\mathrm{mon}}$ in
the SMaA projection~\eq{SMaA} on the lattices $4\cdot
L^3$\footnote{These lattices correspond to finite temperature field
theory, $T = 1/4a$, $a$ is the lattice spacing.}, for $L =
8,10,12,14,16$. We extrapolate the value of the monopole condensate
to the infinite spatial volume ($L \to \infty$) since near the
deconfinement phase transition there are strong finite volume
effects.  We also impose the so--called $C$--periodic boundary
conditions in space for the gauge fields, since the
periodic boundary conditions are forbidden due to the Gauss law:
we input a magnetic charge into the finite box. The $C$--periodic
boundary conditions for the nonabelian
gauge fields \cite{Wie92} correspond to anti--periodic
spatial boundary conditions for abelian fields.
In the case of $SU(2)$ gauge group the
$C$--periodic boundary conditions are almost trivial:  on the
boundary we have $U_{x,\mu} \to \Omega^+ U_{x,\mu}
\Omega,\,\,\Omega = i \sigma_2$.  Note, that the gauge--fixing
functionals for MaA~\eq{MaA} and for SMaA~\eq{SMaA} gauges are
invariant under this transformation.

The effective potential $V(\Phi)$ for the monopole field is defined
via the probability distribution of the operator
$\Phi_{\mathrm{mon}}$, Ref.~\cite{Wiese,OurPapers}:
\beqn
V(\Phi) = -ln(<\delta (\Phi - \Phi_{\mathrm{mon}}(x))>)\,.
\label{V}
\eeqn
We calculated numerically this potential by the Monte-Carlo method. We
generate the gauge fields by the standard heat bath method taking 2000
update sweeps to thermalize the system at each value of $\beta$. The
number of the gauge fixing iterations is defined by the standard
condition~\cite{Pou97}: the iterations are stopped when the matrix of the
gauge transformation $\Omega (x)$ becomes close to the unit matrix: ${\rm
max}_x \{ 1- {\frac 12} Tr \,\Omega (x) \} \le 10^{-5}$.  We check that
more accurate fixing of the SMaA gauge~\eq{SMaA} does not change our
results. To calculate the probability distribution for each value of
$\beta$ at the lattice of definite size, we use 100 gauge field
configurations separated by 300 Monte Carlo sweeps. Then for each field
configuration we calculate the value of the monopole creation
operator~\eq{Ux2} at 20 randomly chosen lattice points.

In Fig.~1 we show the effective potential~\eq{V} for confinement
phase, the lattice is $4\cdot12^3$. This potential
corresponds to the Higgs type potential\footnote{In
Fig.~1 the only right half of the potential is shown due to
positivity of the monopole operator~\eq{Ux2}, see
Refs.~\cite{OurPapers} for a discussion.}.  The value of the
monopole field, $\Phi_c$ at the minimum of the minimum is
equivalent to the value of the monopole condensate.

The minimum of the potential, $\Phi_c$, {\it vs.}  inverse lattice
size, $1 \slash L$, is shown in Fig.~2 for two values of $\beta$.
We fit the data for $\Phi_c(L)$ by the formula $\Phi_c = A L^\alpha
+ \Phi_c^{inf}$, where $A$, $\alpha$ and $\Phi_c^{inf}$ are the
fitting parameters.  We find that the best fit gives $\alpha = -1$
within the statistical errors.

Fig.~3 shows the value of the monopole condensate, extrapolated to
the infinite spatial volume, $\Phi_c^{inf}$.  We conclude from Fig.~3
that in the SMaA projection the infinite--volume condensate
$\Phi_c^{inf}$ vanishes at the point of the phase
transition\footnote{It is well known that for $\beta=\beta_c \approx
2.3$ ($\beta > \beta_c$) the $SU(2)$ gauge field is in the
confinement (deconfinement) phase on the lattice $L^3\times 4$},
$\beta=\beta_c$.  This in the SMaA projection the confining vacuum of
$SU(2)$ gluodynamics behaves as a dual superconductor.

We are grateful to U.--J.~Wiese and to E.--M.~Ilgenfritz for
interesting discussions and to E.--M.~Ilgenfritz, H.~Mar\-kum,
M.~M\"uller-Preussker and S.~Thurner for informing us about their
results prior to publication.  M.N.Ch.  acknowledges the kind
hospitality of the Theory Division of the Max-Planck-Institute for
Physics, Werner Heisenberg Institute, M.I.P.  feel much obliged for
the kind hospitality extended to him by the staff of Centro de
F\'\i sica das Interac\c c\~oes Fundamentais, Edif\'\i cio
Ci\^encia, Instituto Superior T\'ecnico (Lisboa) where a part of
this work has been done.  This work was partially supported by the
grants INTAS-96-370, INTAS-RFBR-95-0681, RFBR-96-02-17230a,
RFBR-97-02-17491a and RFBR-96-15-96740. The work of M.N.Ch. was
supported by the INTAS Grant 96-0457 within the research program of
the International Center for Fundamental Physics in Moscow.

\newpage

\section*{Figures}

\begin{figure}[!htb]
\centerline{\epsfxsize=.60\textwidth\epsffile{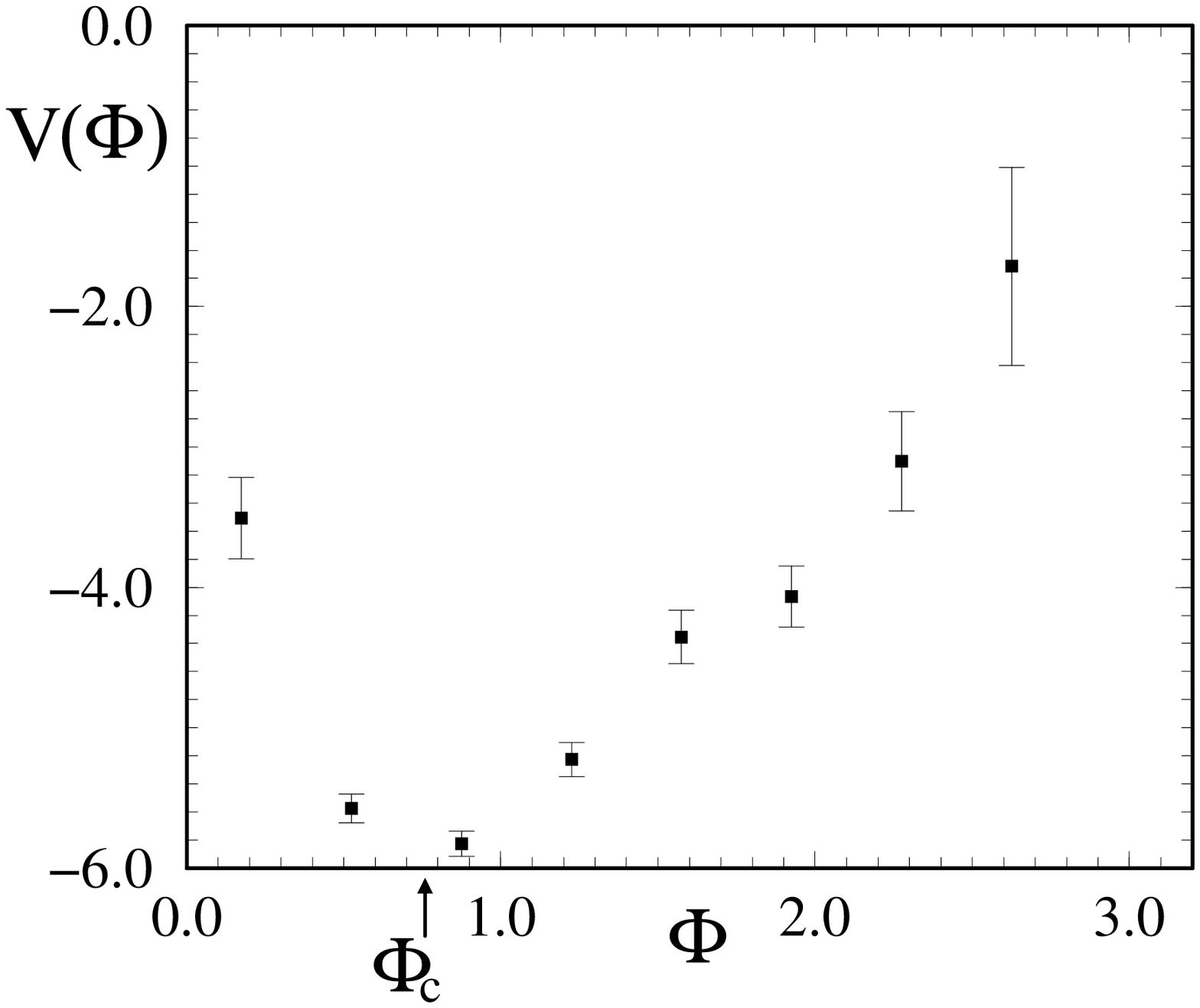}}
\vspace{0.2cm}
\vspace{0.7cm}
\caption{Effective potential $V(\Phi)$, eq.\eq{V},
for confinement phase, $\beta=1.5$.}
\end{figure}

\newpage

\begin{figure}[!htb]
\centerline{\epsfxsize=.60\textwidth\epsfbox{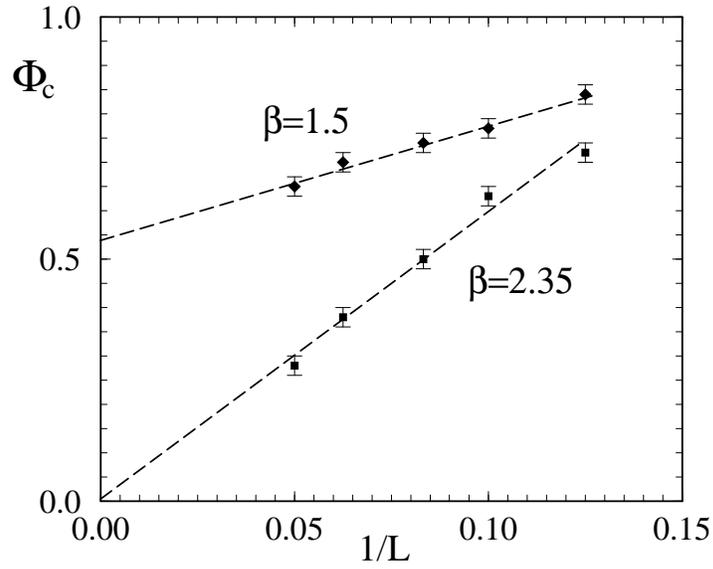}}
\vspace{0.1cm}
\caption{Finite volume monopole condensate, $\Phi_c$, {\it vs.} inverse
spatial size of the lattice, $1 \slash L$, at $\beta=1.5$ and
$\beta=2.35$.}
\end{figure}

\begin{figure}[!htb]
\centerline{\epsfxsize=.58\textwidth\epsfbox{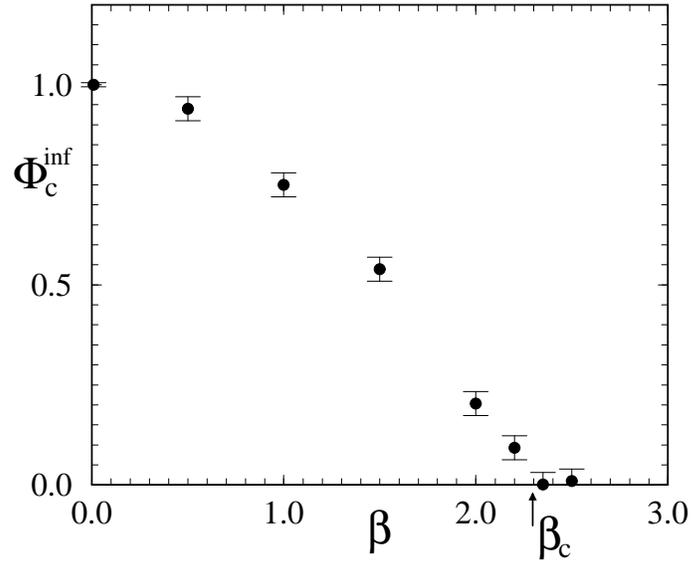}}
\vspace{0.1cm}
\caption{Monopole condensate extrapolated to the
 infinite volume, $\Phi_c^{\mathrm{inf}}$,
{\it vs.} $\beta$. The phase transition is at
$\beta=\beta_c \approx 2.3$.}
\end{figure}

\end{document}